\documentclass[aps,prd,showpacs,superscriptaddress,nofootinbib,floatfix,10pt]{revtex4-2}
\usepackage[utf8]{inputenc}
\usepackage{color}
\usepackage{amsfonts,amsmath,amssymb,leftindex}
\usepackage{graphicx}
\usepackage{epstopdf}
\usepackage{longtable,booktabs}
\usepackage{bm}
\usepackage[colorlinks=true,linkcolor=blue,citecolor=blue]{hyperref}
\usepackage{geometry}
\usepackage{url}
\usepackage{mathtools}
\usepackage{subfigure}
\usepackage{multirow}
\usepackage{diagbox}
\usepackage{afterpage}
\usepackage{placeins}
\usepackage{float}
\usepackage{mathrsfs}
\usepackage{setspace}
\usepackage{orcidlink}
\allowdisplaybreaks[1]

\renewcommand{\arraystretch}{1.5}

\newcommand{\itp}{\affiliation{CAS Key Laboratory of Theoretical Physics, Institute of Theoretical Physics,\\ Chinese Academy of Sciences, Beijing 100190, China}}

\newcommand{\qfnu}{\affiliation{College of Physics and Engineering, Qufu Normal University, Qufu 273165, China}}

\newcommand{\chep}{\affiliation{Center for High Energy Physics, Peking University, Beijing 100871, China}}

\begin{document}

\title{Constrain the \texorpdfstring{$\chi_{cJ}\to D^{(*)}\bar{D}^{(*)} $}{D} effective couplings via the \texorpdfstring{$X(3872)\to \pi^0\chi_{cJ}$}{X(3872)->chicJpi0} decays}

\author{Zhao-Sai Jia\orcidlink{0000-0002-7133-189X}}\qfnu \itp
\author{Gang Li\orcidlink{0000-0002-5227-8296}}\email{gli@qfnu.edu.cn} \qfnu 
\author{Zhen-Hua Zhang\orcidlink{0000-0001-6072-5378}}\email{zhhzhang@pku.edu.cn} \chep

\begin{abstract}
The hidden-charm decays serve as irreplaceable platforms for probing the structures of charmonium-like states, such as $X(3872)$, $Y(4260)$, $Z_c(3900)$, and their heavy-quark-symmetry partners. In the hadronic molecular scenario, these hidden-charm decays are denominated by intermediate meson loops (IMLs), and the couplings of $\chi_{cJ}\to D^{(\ast)}\bar{D}^{(\ast)}$ are building blocks of the amplitudes for the pionic and radiative transitions of the charmonium-like states to the $\chi_{cJ}$ and $h_c$ states, e.g., $X(3872)\to \pi^0\chi_{cJ},\,\pi\pi\chi_{cJ},\,\gamma\chi_{cJ}$ and $Y(4260)\to \pi^0 h_c,\,\eta h_c$. These couplings can not be extracted from the partial decay widths of the $\chi_{cJ}$ directly and only have estimated values from the vector meson dominance (VMD) model. Utilizing the recent precise determination of the pole position and the isospin breaking properties of the $X(3872)$, we give an estimation on the upper bounds of the absolute values of the $\chi_{cJ}\to D^{(\ast)}\bar{D}^{(\ast)}$ couplings. Our results show that the VMD model may over estimate the $\chi_{cJ}\to D^{(\ast)}\bar{D}^{(\ast)}$ couplings considering the $X(3872)$ as a $D\bar{D}^{*}$ hadronic molecule with a binding energy about tens of keV. These upper limits can be used and tested in other hidden-charm transitions of the charmonium-like states to the $\chi_{cJ}$ and $h_c$.
\end{abstract}

\date{\today}

\maketitle

\section{Introduction}\label{sec:Introduction}
Over the past two decades, numerous candidates of exotic hadrons beyond the meson and baryon configurations in the conventional quark model have been observed in the charmonium energy region. These states are referred to as charmonium-like states or $XYZ$ states. The most well-known
charmonium-like states are, for example, $X(3872),\,Y(4260),$ and $Z_c(3900)$. Extensive efforts, both experimental and theoretical, have been made to explore the internal structures of the $XYZ$ states (for recent reviews, see Refs.~\cite{Hosaka:2016pey,Esposito:2016noz,Guo:2017jvc,Olsen:2017bmm,Karliner:2017qhf,Kalashnikova:2018vkv,Brambilla:2019esw,Meng:2022ozq,Liu:2024uxn,Chen:2024eaq}). In these studies, the $XYZ$ states are interpreted as conventional charmonium states, compact tetraquarks, kinematic singularities, and hadronic molecules. Decays of $XYZ$ states to $\chi_{cJ}$ and $h_c$ states with the emission of pions or photons have been widely proposed and studied as crucial probes to distinguish these different interpretations~\cite{Dubynskiy:2007tj,Fleming:2008yn,Li:2013yla,Mehen:2015efa,BESIII:2019esk,Belle:2019jeu,Zhou:2019swr,Wu:2021udi,Wang:2023sii,BESIII:2023eeb,Achasov:2024ezv,BESIII:2024ilt,Achasov:2024anu,Liu:2024ogo,Liu:2025sjz}. For different configurations of the charmonium-like states, these decays can involve different mechanisms, resulting in significantly different decay widths. In the hadronic molecular scenario, the hidden-charm transitions of the $XYZ$ states proceed predominantly through intermediate meson loops (IMLs) of open-charm particles~\cite{Guo:2010ak,Guo:2017jvc}. In this framework, the couplings of the final-state charmonium to charmed meson pairs ($g_1$ for $\chi_{cJ}/h_c\to D^{(*)}\bar{D}^{(*)}$) are essential components of the decay amplitude. However, the coupling $g_1$ cannot be directly extracted from their partial widths and only have an estimated value from the vector meson dominance (VMD) model~\cite{Colangelo:2003sa,Deandrea:2003pv}, $g_1^2\approx m_{\chi_{c0}}/(6f_{\chi_{c0}}^2)$, where $m_{\chi_{c0}}$ is the mass of $\chi_{c0}$, and the value of the $\chi_{c0}$ decay constant $f_{\chi_{c0}}$ is calculated to be about $300\sim 500$ MeV in the QCD sum rules~\cite{Novikov:1977dq,Veliev:2010gb,Colangelo:2002mj}. Such estimation could bring large uncertainties in the partial decay widths of the $XYZ$ states~\cite{Fleming:2008yn,Mehen:2015efa}. Recently, the properties of the $X(3872)$ states, including its pole positions, compositeness, and isospin breaking effects, have been determined precisely in Refs.~\cite{Dias:2024zfh,Ji:2025hjw} from the newly updated experimental data from the BESIII~\cite{BESIII:2023hml} and LHCb~\cite{LHCb:2020fvo,LHCb:2022jez} Collaborations. This offers great opportunities to put limitations on the $g_1^2$ from the $X(3872)\to \pi^0\chi_{cJ}$ isospin-breaking decays.

The $X(3872)$, also known as $\chi_{c1}(3872)$~\cite{ParticleDataGroup:2024cfk}, was first discovered in 2003 by the Belle Collaboration~\cite{Belle:2003nnu} in the $\pi^+\pi^-J/\psi$ invariant mass spectrum from $B^\pm \to (X(3872)\to \pi^+\pi^-J/\psi)K^\pm$ produced in the $e^+e^-$ collisions, with a mass of $3872.0\pm0.6\pm0.5$ MeV, a width less than $2.3$ MeV, and quantum numbers $I^G(J^{PC})=0^+(1^{++})$~\cite{ParticleDataGroup:2024cfk,Belle:2011vlx,BaBar:2004cah,LHCb:2013kgk,LHCb:2015jfc}.
The most striking characteristic of the $X(3872)$ is that its mass coincides exactly with the $D^0\bar{D}^{*0}$ threshold, with a difference (hereinafter called binding energy) $BE_n=m_{D^0}+m_{D^{*0}}-m_{X(3872)}=(0.01 \pm 0.14)$ MeV~\cite{LHCb:2020xds} and $(0.12 \pm 0.13)$ MeV~\cite{LHCb:2020fvo}, where $m_{D^0}=(1864.84 \pm 0.05)$ MeV, $m_{D^{*0}}=(2006.85 \pm 0.15)$ MeV~\cite{ParticleDataGroup:2020ssz}. The newly updated study~\cite{Ji:2025hjw} determined the pole position of the $X(3872)$ to be $E_X=-53_{-24}^{+9}-i 34_{-12}^{+2}$~keV relative to the $D^0\bar{D}^{*0}$ threshold. Despite the closeness to the threshold, the $X(3872)$ exhibits strong coupling to the $D^0\bar{D}^{*0}$ channel, as evidenced by the large branching fraction to the $D^0\bar{D}^{*0}$ and $D^0\bar{D}^0 \pi^0$ channels~\cite{ParticleDataGroup:2024cfk}. Because of these properties, the $X(3872)$ is widely hypothesized to be a $D\bar{D}^{*}$ hadronic molecule. The molecular component of the $X(3872)$ can be quantified by its compositeness (see Ref.~\cite{Esposito:2025hlp} for a recent review). In a newly updated detailed study~\cite{Ji:2025hjw}, the compositeness of the $X(3872)$ is determined to be $\tilde{X}_{A}>0.99$, strongly support the dominance of the molecular component.

Another important feature of the $X(3872)$ is the 
significant isospin-breaking in its decays, such as $X(3872) \to \pi^+\pi^-J/\psi$, $X(3872) \to \pi^0 \chi_{cJ}$, and $X(3872)\to\pi^0\pi^+\pi^-$~\cite{Belle:2005lfc,BaBar:2010wfc,BESIII:2019qvy,BESIII:2019esk,Belle:2019jeu,Suzuki:2005ha,DELPHI:2003cnx,Meng:2007cx,Dubynskiy:2007tj,Fleming:2008yn,Gamermann:2009fv,Gamermann:2009uq,Terasaki:2009in,Karliner:2010sz,Li:2012cs,Mehen:2015efa,Achasov:2019cfe,Zhou:2019swr,Wu:2021udi,Meng:2021kmi,LHCb:2022jez,Wang:2023sii,Achasov:2019wvw,Belle:2022puc,Dias:2024zfh}. The isospin-breaking decay $X(3872) \to \pi^+\pi^-J/\psi$, which proceeds predominantly via the $\rho^0J/\psi$ intermediate state, has a branching ratio comparable to that of the isospin-conserving decay $X(3872) \to \omega J/\psi$,
\begin{align}
    \frac{\mathrm{Br}(X(3872) \rightarrow \omega J / \psi)}{\mathrm{Br}(X(3872) \rightarrow \pi^{+} \pi^{-} J / \psi)}=1.1 \pm 0.4.
\end{align}
Given the limited phase space for $X(3872) \to \omega J/\psi$ decay, the latest result from the LHCb Collaboration for the amplitude ratio $R_X$ of $X(3872) \to \rho^0 J/\psi$ and $X(3872) \to \omega J/\psi$~\cite{LHCb:2022jez} is
\begin{align}
    R_X=\left|\frac{\mathcal{M}(X(3872) \rightarrow \rho^0 J / \psi)}{\mathcal{M}(X(3872) \rightarrow \omega J / \psi)}\right|=0.29 \pm 0.04,
    \label{eq_RX}
\end{align}
which is about one order of magnitude larger than that of conventional charmonium. The value of $R_X$ is recently determined more precisely to be $R_X=0.28\pm 0.04$~\cite{Ji:2025hjw} and $R_X=0.26\pm 0.03$~\cite{Dubynskiy:2007tj} using dispersive formalisms with and without the contribution of isovector $W_{c1}^0$~\cite{Zhang:2024fxy}. Such a significant isospin-breaking effect may originate from the mass discrepancy between charged and neutral charm mesons. Studies~\cite{Aceti:2012cb,Aceti:2012qd} have highlighted the necessity of incorporating charged $D$ mesons in the calculations of decays $X(3872) \to \gamma J/\psi$, $\pi^+ \pi^- J/\psi$,
and $\pi^+\pi^-\pi^0 J/\psi$.

Prior to the experimental study from the BESIII~\cite{BESIII:2019esk} and Belle~\cite{Belle:2019jeu} Collaborations, the $X(3872) \to \pi^0\chi_{cJ}$ decay had been proposed in Refs.~\cite{Dubynskiy:2007tj,Fleming:2008yn,Mehen:2015efa} assuming the $X(3872)$ being a charmonium or molecular state. Subsequently, this decay mode was further investigated in various theoretical frameworks in, e.g., Refs.~\cite{Zhou:2019swr,Wu:2021udi,Wang:2023sii,Achasov:2024ezv}. In Ref.~\cite{Fleming:2008yn}, the $X(3872)$ was interpreted as a loosely bound state of neutral charmed mesons $D^0\bar{D}^{*0}$, and its decay into $\pi^0 \chi_{cJ}$ mainly proceeds via a neutral charmed meson loop. The studies in Ref.~\cite{Mehen:2015efa} pointed out that the $D^+D^{*-}$ components in the $X(3872)$ and thus the charged meson loop also plays an important role in these isospin-breaking decays, otherwise the partial width of $X(3872)\to\pi^0\chi_{cJ}$ would exceed the total width of the $X(3872)$~\cite{ParticleDataGroup:2024cfk}. Therefore, the partial width of $X(3872)\to\pi^0\chi_{cJ}$ are sensitive to the relative amount of the $D^0\bar{D}^{*0}$ and $D^+D^{*-}$ components in the $X(3872)$ wave function, which can be quantified by the mixing angle $\theta$~\cite{Cai:2025inq}, 
 \begin{align}
    |X(3872)\rangle = \frac{1}{\sqrt{2}} \cos{\theta}(|D^0\bar{D}^{*0}\rangle+|D^{*0}\bar{D}^0\rangle)+\frac{1}{\sqrt{2}} \sin{\theta}(|D^+D^{*-}\rangle+|D^-D^{*+}\rangle),
\label{eq_X}
\end{align}
where the charge conjugation conventions $D^* \overset{C}{\rightarrow} \bar{D}^*$ and $D \overset{C}{\rightarrow} \bar{D}$ are implemented. The mixing angle $\theta$ can be determined from the isospin-breaking decay properties of the $X(3872)$~\cite{Zhang:2024fxy}. Due to the lack of precise experimental data, the mixing angle was estimated to be $\theta=0.37\pm0.04$ through the maximum cancellation between the charged and neutral meson loops in Ref.~\cite{Mehen:2015efa}. The partial widths for different mixing angles of the isospin-breaking decays $X(3872) \to \pi^+\pi^-J/\psi$ and $X(3872) \to\pi^0\chi_{cJ}$ were also investigated in Ref.~\cite{Wu:2021udi}. All of these works utilized the VMD model-estimated value of the $\chi_{cJ}\to D^{(*)}\bar{D}^{(*)}$ coupling $g_1$ in the IMLs, which brings further uncertainties to the final results. With the input from the precisely determined isospin-breaking amplitude ratio $R_X$, the mixing angle $\theta$ can be settled specifically, and the upper limit of the branching ratio of the $X(3872)\to \pi^0\chi_{cJ}$ can be used to put an upper bound on the $g_1^2$.

In this work, we take the decay $X(3872) \to \pi^0\chi_{c0}(1P)$ serves as a pivotal probe to investigate the coupling between $\chi_{cJ}$ and charmed mesons. We first utilize the $R_X$ extracted from the LHCb and BESIII data~\cite{Ji:2025hjw} to determine the proportion of neutral and charged constituents in $X(3872)$, and then put an upper limit on $g_1^2$ through the branch ratio of the $X(3872) \to \pi^0\chi_{c0}(1P)$~\cite{ParticleDataGroup:2024cfk} decay. Finally, for a consistency check, the upper limit of $g_1^2$ determined from the $X(3872) \to \pi^0\chi_{c0}(1P)$ decay is applied to the $X(3872) \to \pi^0\chi_{c1,2}(1P)$ decays to give the upper bounds of their partial widths.

The structure of the paper is as follows. In Sec.~\ref{sec:Framework}, we introduce the effective Lagrangians and Feynman diagrams for $X(3872) \to \pi^0\chi_{cJ}$. The numerical results of the upper bounds of $g_1^2$ and the $X(3872) \to \pi^0\chi_{c1,2}(1P)$ branching ratios are presented in Sec.~\ref{sec:Numerical Results}, and a brief summary is given in Sec.~\ref{sec:Summary}.

\section{Framework}\label{sec:Framework}
\begin{figure*}[tb]
    \centering
    \includegraphics[width=0.95\linewidth]{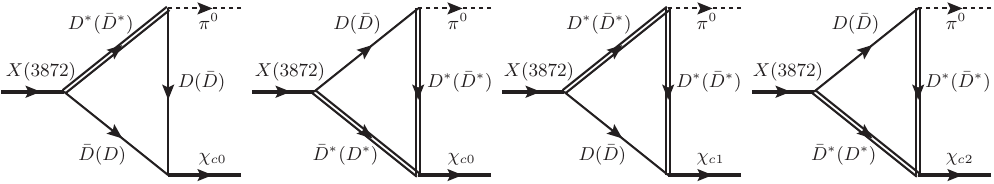}
      \caption{Feynman diagrams for $X(3872) \to \pi^0\chi_{cJ}$. The double lines represent vector mesons ($D^*, \bar{D}^*$), the single solid lines represent scalar mesons ($D, \bar{D}$), and the dashed lines represent the scalar meson $\pi^0$.}
    \label{fig_Feynmandiagrams}
\end{figure*}
The Feynman diagrams for $X(3872) \to \pi^0\chi_{cJ}(1P)$ are shown in Fig.~\ref{fig_Feynmandiagrams}. We assume the $X(3872)$ is an $S$-wave molecular state with $J^{PC}=1^{++}$, represented as the superposition $D^0\bar{D}^{*0}+\mathrm{c.c}$ and $D^-D^{*+}+\mathrm{c.c}$ given in Eq.~\eqref{eq_X}. Correspondingly, the effective Lagrangian describing the coupling of the $X(3872)$ to $D^* \bar{D}$ can be written as 
\begin{align}
\mathcal{L}_{XDD^*}=\frac{g_n}{\sqrt{2}}X^{i \dagger}(D^{*0i}\bar{D}^0 +D^0\bar{D}^{*0i}) + \frac{g_c}{\sqrt{2}}X^{i \dagger}(D^{*+i}D^- +D^+D^{*-i})+\text{H.c.}.
\label{eq_L_XDD}
\end{align}
The coupling constants of the $X(3872)$ to neutral and charged charmed mesons, $g_n$ and $g_c$, can be extracted from the residues of the $D^0\bar{D}^{*0}-D^{+}D^{*-}$ coupled-channel scattering $T$-matrix at the $X(3872)$ pole position~\cite{Albaladejo:2015dsa} as,
\begin{align}
    g_n=4\sqrt{\frac{\pi \gamma_n}{\mu_n^2}} \cos \theta, \quad g_c=4\sqrt{\frac{\pi \gamma_c}{\mu_c^2}} \sin \theta,
    \label{eq_gngc}
\end{align}
where $\gamma_n=\sqrt{2\mu_n BE_n}$, $\gamma_c=\sqrt{2\mu_c BE_c}$ with $BE_n=m_{D^{*0}}+m_{D^0}-m_{X(3872)}$, $BE_c=m_{D^{*+}}+m_{D^-}-m_{X(3872)}=\Delta+BE_n$ being the binding energies of the $X(3872)$ relative to the thresholds of the neutral and charged channels, respectively. 
Here, $\Delta=m_{D^{*+}}+m_{D^-}-m_{D^{*0}}-m_{D^0}$ is the gap between the thresholds of $D^0\bar{D}^{*0}$ and $D^+\bar{D}^{*-}$, and $\mu_n$ and $\mu_c$ are the reduced masses of $D^0\bar{D}^{*0}$ and $D^+D^{*-}$, respectively. These coupling constants can be related to the amplitude ratio $R_X$ for $X(3872) \to \rho^0 J/\psi$ and $X(3872) \to \omega J/\psi$ by~\cite{LHCb:2022jez,Zhang:2024fxy} ,
\begin{align}
    R_X=\left|\frac{1-R_{c/n}}{1+R_{c/n}}\right|, \quad R_{c/n} \equiv \frac{g_c}{g_n}=\left(\frac{\mu_n}{\mu_c}\right)^{\frac{3}{4}}\left(1+\frac{\Delta}{BE_n}\right)^{\frac{1}{4}}\tan \theta.
    \label{eq_RXgcn}
\end{align}
 Therefore, $R_X$ can be used as an input to constrain the mixing angle $\theta$. There are two solutions of $R_{c/n}$, $R_{c/n}=0.56\pm0.05$ and $R_{c/n}=1.78^{+0.16}_{-0.15}$. The later one means the coupling of the $X(3872)$ to $D^+D^{*-}$ is stronger than to $D^0\bar{D}^{*0}$. However, this scenario is not supported by experimental evidence, which indicates the opposite hierarchy. Therefore, the second solution should be excluded.

The leading-order effective Lagrangian for the $D^{(*)}D^*\pi$ coupling is given by the heavy hadron chiral perturbation theory (HH$\chi$PT) as~\cite{Mehen:2015efa,Fleming:2008yn}
\begin{align}
    &\mathcal{L}_{D^{(*)}D^*\phi} =-g \text{Tr}[H_a^\dagger H_b \vec \sigma \cdot \vec A_{ba}]+g \text{Tr}[\bar{H}_a^\dagger \vec \sigma \cdot \vec A_{ab} \bar{H}_b],
    \label{eq_L_DDphi}
\end{align}
where $\vec \sigma$ denote the Pauli matrices and $a$ is the light flavor index, the charmed mesons are given by the two-component notation~\cite{Hu:2005gf} as $H_a = \vec V_a \cdot \vec \sigma + P_a$ with $P=(D^-, \bar{D}^0)$ and $V=(D^{*-}, \bar{D}^{*0})$ being the pseudoscalar and vector heavy mesons, respectively, and the field for the antimesons is $\bar{H}_a=-\vec{\bar{V}}_a \cdot \vec{\sigma} +\bar{P}_a$ with $\bar{P}=(D^+, D^0)$ and $\bar{V}=(D^{*+}, D^{*0})$. The field $\vec{A}_{ab}=-\vec \bigtriangledown{\Phi}_{ab}/f_{\pi}+\cdots$ is the axial current in the chiral perturbation theory ($\chi$PT) and couples to the heavy mesons with the axial coupling $g=0.54$~\cite{Mehen:2015efa}, where $f_{\pi}=130~\rm{MeV}$ is the pion decay constant, and the $\Phi$ field contains the Goldstone bosons as components,
\begin{align}
\Phi=\left( 
\begin{array}{cc}
\frac{1}{\sqrt{2}}\pi^{0} & \pi^{+} \\
\pi^{-} & -\frac{1}{\sqrt{2}}\pi^{0}
\end{array}\right).
\end{align}

The Lagrangian couples the $\chi_{cJ}$ to the $D^{(*)}$ mesons reads
\begin{align}
& \mathcal{L}_{\chi D^{(*)}D^{(*)}}= i \frac{g_1}{2} \text{Tr} \left[ \chi^{i \dagger} H_a \sigma^i \bar{H}_a \right]+ \text{H.c.},
\label{eq.L_chiDD}
\end{align}
where the $\chi_{cJ}$ field is expressed as~\cite{Fleming:2008yn}
\begin{align}
    \chi^i=\sigma^j \chi^{ij}=\sigma^j \left( \chi_{c2}^{ij} +\frac{1}{\sqrt{2}} \varepsilon ^{ijk} \chi_{c1}^k +\frac{\delta^{ij}}{\sqrt{3}} \chi_{c0} \right)+h_c^i,
\end{align}
with the unknown coupling of the $\chi_{cJ}$ to charmed mesons $g_1$. In Ref.~\cite{Mehen:2015efa}, the coupling constant was estimated in the VMD model as $g_1^2 \approx m_{\chi_{c0}}/(6f_{\chi_{c0}}^2)=1/(457~\mathrm{MeV})$ with $f_{\chi_{c0}}=510$ MeV from the QCD sum rule calculation~\cite{Novikov:1977dq}.
In this study, we will first determine the range of the mixing angle $\theta$ using the input value $R_X=0.28(4)$ from Ref.~\cite{Ji:2025hjw}, and then utilize the upper bounds $\Gamma(X(3872) \to \pi^0\chi_{c0}(1P))/\Gamma_{X(3872)} < 16\%$~\cite{ParticleDataGroup:2024cfk} and $\Gamma_{X(3872)}\lesssim131$ keV~\cite{Mehen:2015efa} to constrain the coupling constant square $g_1^2$. The upper bound of $\Gamma_{X(3872)}$ in Ref.~\cite{Mehen:2015efa} was derived from the theoretical calculation of  $\Gamma(X(3872)\to D^0\bar{D}^0\pi^0)$ within the molecular framework, combined with the experimentally measured lower bound of the $X(3872)\to D^0\bar{D}^0\pi^0$ branching ratio~\cite{ParticleDataGroup:2024cfk}.
\begin{table}[b]
    \renewcommand\arraystretch{2}
    \centering
    \caption{All the possible combinations of the intermediate charmed mesons for the diagrams in Fig.~\ref{fig_Feynmandiagrams}. The first particle in each square bracket denotes top left intermediate charmed meson in the corresponding diagram, and the other intermediate charmed mesons in the same diagram are listed in the square bracket in counterclockwise order along the loop.}
    \begin{tabular}{c|c}
    \hline\hline
    \multirow{2}{*}{
    $X(3872) \to \pi^0\chi_{c0}$} &$[D^{*0}, \bar{D}^0,D^0]$, $[\bar{D}^{*0}, D^0, \bar{D}^0]$, $[D^{*+}, D^-,D^+]$, $[D^{*-}, D^+, D^-]$,  \\
    &$[\bar{D}^0, D^{*0}, \bar{D}^{*0}]$, $[D^0, \bar{D}^{*0}, D^{*0}]$, $[D^-, D^{*+}, D^{*-}]$, $[D^+, D^{*-}, D^{*+}]$\\
    \hline
    $X(3872) \to \pi^0\chi_{c1}$ &$[D^{*0}, \bar{D}^0,D^{*0}]$, $[\bar{D}^{*0}, D^0, \bar{D}^{*0}]$,$[D^{*+}, D^-,D^{*+}]$, $[D^{*-}, D^+, D^{*-}]$ \\
    \hline
    $X(3872) \to \pi^0\chi_{c2}$ &$[\bar{D}^0, D^{*0}, \bar{D}^{*0}]$, $[D^0, \bar{D}^{*0}, D^{*0}]$,$[D^-, D^{*+}, D^{*-}]$, $[D^+, D^{*-}, D^{*+}]$ \\
    \hline\hline
    \end{tabular}
    \label{Tab:loop particle}
\end{table}
Based on the Lagrangians given above, the loop transition amplitudes in Fig.~\ref{fig_Feynmandiagrams} can be expressed in a general form as follows,
\begin{align}
 \mathcal{M}(X(3872) \to \pi \chi_{cJ})=&\,\sqrt{m_{X(3872)} m_{\chi_{cJ}}} m_1 m_2 m_3 V_1 V_2 V_3 \times I[m_1,m_2,m_3],
\end{align}
where $V_l$ ($l=1, 2, 3$) represent the vertexes of the charmed mesons in the loop coupling to the initial $X(3872)$, final $\chi_{cJ}$ and $\pi^0$, respectively, and the prefactor $\sqrt{m_{X(3872)} m_{\chi_{cJ}}} m_1 m_2 m_3$ accounts for the relativistic normalization. The $I[m_1, m_2, m_3]$ represents the scalar 3-point integral, where $m_k~(k=1, 2, 3)$ represents the mass of $k$-th particle of each combination of the intermediate states in Table~\ref{Tab:loop particle} for different processes. The first particle in each square bracket denotes the top left intermediate charmed meson in the corresponding triangle diagram, and the other intermediate charmed mesons in the same diagram are listed in the square bracket in counterclockwise order along the loop. The scalar 3-point loop integral $I[m_1,m_2,m_3]$ is ultraviolet (UV) convergent and can be worked out as~\cite{Guo:2010ak}
\begin{align}
I[m_1,m_2,m_3]&=i\int \frac{d^4l}{(2 \pi)^4} \frac{1}{[l^2-m_1^2+i\epsilon][(p-l)^2-m_2^2+i\epsilon][(l-q)^2-m_3^2+i\epsilon]}\nonumber \\
&\simeq \frac{\mu_{12} \mu_{23}}{16 \pi m_1 m_2 m_3} \frac{1}{\sqrt{a}}\left[\tan ^{-1}\left(\frac{c_2-c_1}{2 \sqrt{a c_1}}\right)+\tan ^{-1}\left(\frac{2 a+c_1-c_2}{2 \sqrt{a\left(c_2-a\right)}}\right)\right],
\label{Eq:3-piont loop_integral}
\end{align}
where $p, l,$ and $q$ are the 4-momenta of the initial $X(3872)$, the top left intermediate charmed meson in the triangle diagram, and the final-state $\pi^0$ in the $X(3872)$ rest frame, respectively, $\mu_{ij}=m_{i} m_{j} /\left(m_{i}+m_{j}\right)$ are the reduced masses, and 
\begin{align}
 a=\left(\frac{\mu_{23}}{m_3}\right)^2 \vec{q}\,^2, \quad c_1=2 \mu_{12} b_{12}, \quad c_2=2 \mu_{23} b_{23}+\frac{\mu_{23}}{m_3} \vec{q}\,^2,
\end{align}
with $b_{12}=m_1+m_2-M$, $b_{23}=m_{2}+m_{3}+q^0-M$, and the magnitude of 3-momentum of the final state in the initial state rest frame 
\begin{align}
    |\vec{q}\,|=\frac{\sqrt{(m_{X(3872)}-(m_{\chi_{cJ}}+m_\pi)^2)(m_{X(3872)}-(m_{\chi_{cJ}}-m_\pi)^2)}}{2m_{X(3872)}}.
\end{align}

 All the combinations in each diagram should be summed to get the final amplitude, and  $X(3872)\to \pi^0\chi_{cJ}$ decay widths are given by
\begin{align}
\Gamma(X(3872)\to \pi^0\chi_{cJ})=\frac{1}{2j+1} \frac{|\vec{q}\,|}{8\pi m^2_{X(3872)}} \sum_{\text{spins}} \left\vert \mathcal{M}\right\vert^2,
\label{Eq.X decay rate}
\end{align}
where $j=1$ is the spin of the initial $X(3872)$,
and there is a sum over all polarizations of the final-state particles.

\section{Numerical Results}\label{sec:Numerical Results}

\begin{figure}[htbp]
      \subfigure[] {
      \includegraphics[scale=0.612]{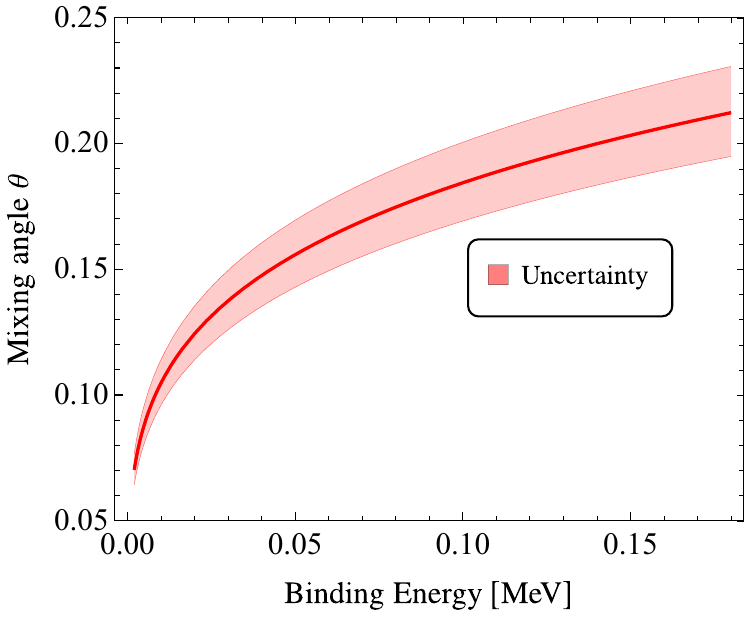}
      \label{fig_thetavsBE}
      } 
      \subfigure[] {
      \includegraphics[scale=0.6]{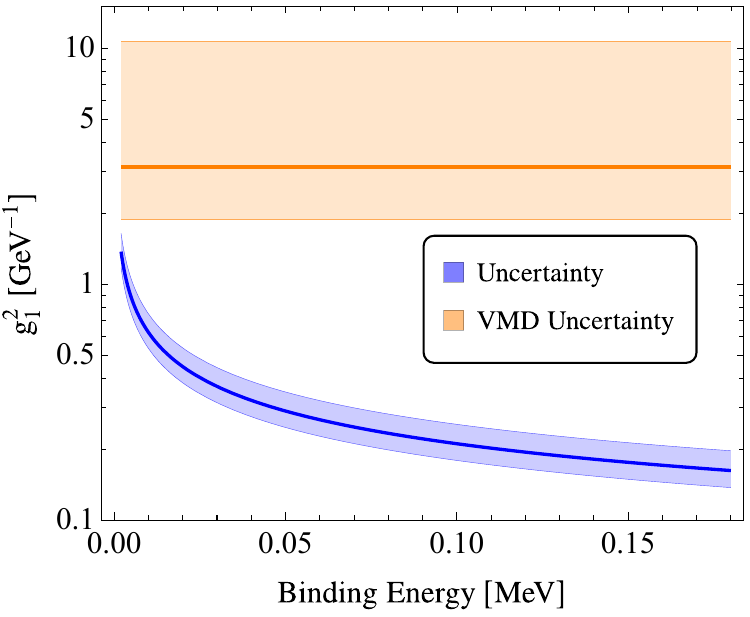}
      \label{fig_g1vsBE}
      } \caption{The mixing angle $\theta$ describing the proportion of neutral and charged constituents in $X(3872)$ (a), and the upper limits of the coupling constant square $g_1^2$ of $\chi_{cJ}$ to charmed mesons (b) as a function of the binding energy $BE_n$. The red and blue bands show the uncertainties of $\theta$ and $g_1^2$ determined from the $X(3872)$ properties propagated from the uncertainty of $R_X$, and the orange band gives the uncertainty of $g_1^2$ from the VMD model propagated from the uncertainty of the $\chi_{c0}$ decay constant $f_{\chi_{c0}}$. }
    \label{fig_thetag1}
\end{figure}

In this section, we present the allowed ranges of the mixing angle $\theta$ and coupling constant square $g_1^2$. The angle $\theta$ characterizes the proportion of neutral and charged constituents within the $X(3872)$, while $g_1^2$ describes the interaction strength between $\chi_{cJ}$ and charmed mesons. These quantities are determined with the binding energy $BE_n$ of the $X(3872)$ varying from 2 keV to 180 keV.

Fig.~\ref{fig_thetavsBE} illustrates the dependence of the mixing angle $\theta$ on $BE_n$ over the range $[2, 180]~\mathrm{keV}$. As shown, $\theta$ increases with increasing $BE_n$. This trend implies a systematic rearrangement of the internal composition of $X(3872)$ as its binding energy changes. Nevertheless, the neutral constituent remains dominant across the explored range of $BE_n \in [2, 180]~\mathrm{keV}$. Taking the central value of $\theta$ at $BE_n = 53~\mathrm{keV}$ as representative and incorporating uncertainties from both $BE_n$ and $R_X$, we obtain the mixing angle:
\begin{align}
\theta=0.16^{+0.07}_{-0.09}.
\end{align}

The variation of the upper limit of the coupling constant square $g_1^2$ determined from the upper bound of $\Gamma(X(3872)\to \pi^0\chi_{c0})$ versus $BE_n\in[2,\,180]$~keV is presented by the blue curve in Fig.~\ref{fig_g1vsBE}. One can see that the $g_1^2$ decreases monotonically as $BE_n$ increases. Taking the central value of $g_1^2$ to be that at $BE_n=53$~keV and considering the uncertainties from $BE_n$ and $R_X$, the upper bound of $g_1^2$ derived from the $X(3872)$ partial width constraints is
\begin{align}
    g_1^2\lesssim0.28^{+1.36}_{-0.14}~\rm{GeV^{-1}}.
    \label{eq:g1sq_X}
\end{align}
The $g_1^2\approx m_{\chi_{c0}}/(6 f_{\chi_{c0}}^2)$ from the VMD model is shown by the orange line in Fig.~\ref{fig_g1vsBE}, with $m_{\chi_{c0}}=3414.71$~MeV~\cite{ParticleDataGroup:2024cfk}. The uncertainty shown by the orange bind mainly comes from the uncertainty of $f_{\chi_{c0}}$ calculated from the QCD sum rules. We have taken the average central value of $f_{\chi_{c0}}=510\pm 40$~MeV in Ref.~\cite{Novikov:1977dq} and $f_{\chi_{c0}}=343\pm 112$~MeV in Ref.~\cite{Veliev:2010gb} to evaluate the central value of $g_1^2$. The VMD model estimation of the $g_1^2$ is
\begin{align}
 g_{1}^2(\mathrm{VMD})=3.13^{+7.54}_{-1.25}~\rm{GeV^{-1}},
\end{align}  
whose central value is one order of magnitude larger than that determined from the $X(3872)$ decay in Eq.~\eqref{eq:g1sq_X}. The lower limit of the $g_1^2$ from the VMD model is still larger than the upper limit of that from the $X(3872)$ decay. Therefore, the VMD model may over estimate the coupling of $\chi_{cJ}\to D^{(*)}\bar{D}^{(*)}$, at least for the $X(3872)$ being a $D\bar{D}^{*}$ bound state with its binding energy $BE_n$ at tens of keV.

\begin{figure}[htbp]
      \subfigure[] {
      \includegraphics[scale=0.61]{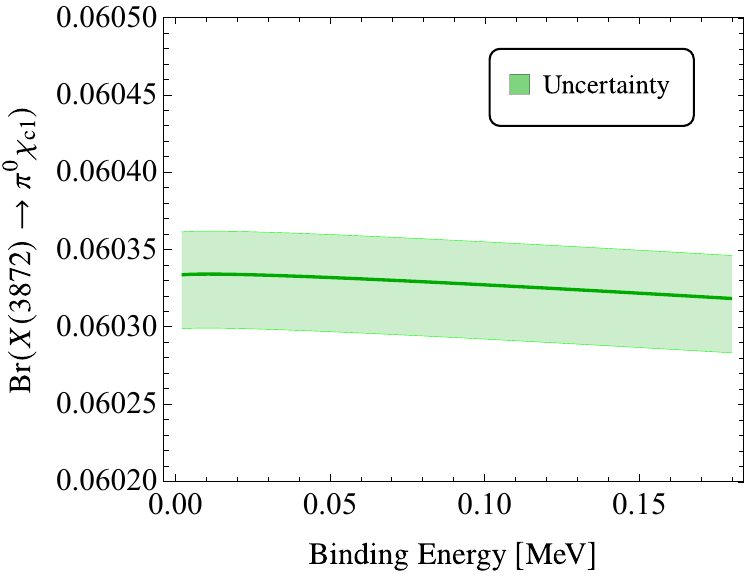}
      \label{fig_Br1X}
      } 
      \subfigure[] {
      \includegraphics[scale=0.6]{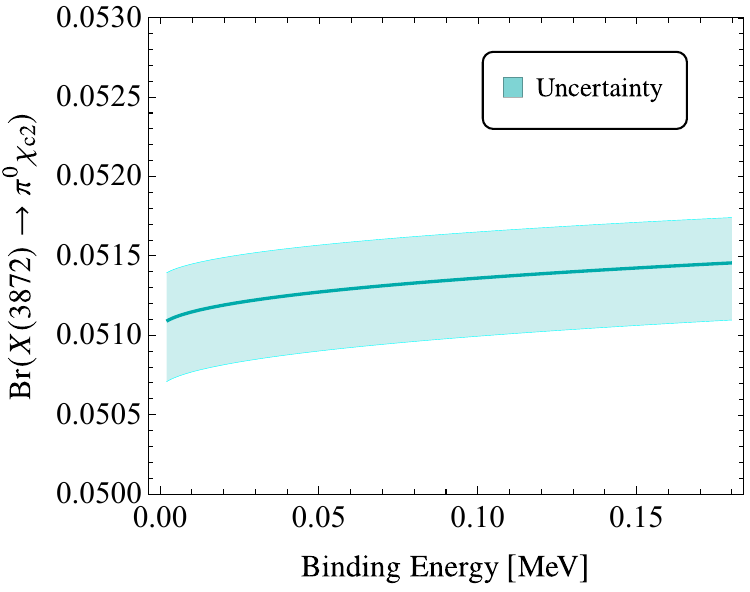}
      \label{fig_Br2X}
      } \caption{Branching ratios of $X(3872) \to \pi^0\chi_{c1}$ (a) and $X(3872) \to \pi^0\chi_{c2}$ (b) as functions of the binding energy $BE_n$ calculated by utilizing the $g_1^2$ determined from the $X(3872) \to \pi^0\chi_{c0}$ decay. The uncertainties are propagated from that of $R_X$.}
      \label{fig_Br12X}
\end{figure}
To check the validity of our result, we calculate the upper bounds of the branching ratios of the 
$X(3872) \to \pi^0 \chi_{c1,2}$ decays using the $g_1^2$ determined from the $X(3872) \to \pi^0 \chi_{c0}$ decay. These branching ratios are shown in Fig.~\ref{fig_Br12X} as functions of the $X(3872)$ binding energy $BE_n$. The upper bounds are $\mathrm{Br}(X(3872)\to \pi^0\chi_{c1})\lesssim 6\%$ and $\mathrm{Br}(X(3872)\to \pi^0\chi_{c2})\lesssim 5\%$,
consistent with the experimental results~\cite{ParticleDataGroup:2024cfk}.

\section{Summary}\label{sec:Summary}

In this work, we employed the amplitude ratio $R_X$ of $X(3872) \to \rho^0 J/\psi$ and $X(3872) \to \omega J/\psi$~\cite{LHCb:2022jez,Zhang:2024fxy} as an input to determine the mixing angle $\theta$ of the $D^0\bar{D}^{*0}$ and $D^+D^{*-}$ components in the $X(3872)$, where $X(3872)$ is assumed to be a $D\bar{D}^*$ hadronic molecule. For the $X(3872)$ binding energy $BE_n\in [2, 180]$ keV, we constrain $\theta=0.16^{+0.07}_{-0.09}$, indicating a dominant $D^0\bar{D}^{*0}$ component in $X(3872)$. Further, we utilize the mixing angle and the upper limit of the branching fraction $\Gamma(X(3872) \to \pi^0\chi_{c0}(1P))/\Gamma_{X(3872)} < 16\%$~\cite{ParticleDataGroup:2024cfk} with $\Gamma_{X(3872)}\lesssim131$ keV~\cite{Mehen:2015efa}, to constrain the upper bound of the coupling constant square $g_1^2$ for $\chi_{cJ} \to D^{(*)}\bar{D}^{(*)}$. The $g_1^2$ is determined to be $g_1^2\lesssim0.28^{+1.36}_{-0.14}~\rm{GeV^{-1}}$, whose central values is about one order-of-magnitude smaller than that from the VMD model estimation.

To check the reliability of our result about $g_1^2$, we compute the partial decay widths of $X(3872) \to \pi^0 \chi_{c1,2}$ using the $g_1^2$ determined from the $X(3872) \to \pi^0 \chi_{c0}$ decay. The results is consistent with the experimental measurement~\cite{ParticleDataGroup:2024cfk}. Our results of $g_1^2$ can be further used and examined in other hidden-charm decay processes of the $XYZ$ states, e.g.,
$X(3872)\to \pi\pi\chi_{cJ}$, $Y(4260)\to \pi h_c$  and $Z_c(3900)/Z_c(4020)\to \pi h_c$.

\section{ACKNOWLEDGMENTS}\label{sec: ACKNOWLEDGMENTS}

This work is partly supported by the National Natural Science Foundation of China under Grant Nos.
12475081, 11835015, and 12047503, the Natural Science
Foundation of Shandong Province under Grant No. ZR2022ZD26, and by the Postdoctoral Fellowship Program of China Postdoctoral
Science Foundation under Grant No. 2025M773427. It is also supported by Taishan
Scholar Project of Shandong Province (Grant No.tsqn202103062).
\bibliography{chicjg1}
\end{document}